\documentclass[12pt]{iopart}

\usepackage{graphicx}
\usepackage{dcolumn}

\newcolumntype{x}{D{.}{.}{0}}
\newcolumntype{y}{D{.}{.}{1}}
\newcolumntype{z}{D{.}{.}{2}}
\newcolumntype{t}{D{.}{.}{4}}
\newcolumntype{u}{D{.}{.}{5}}

\newcommand{\Var}{\ensuremath{\mbox{Var}}}
\newcommand{\Cov}{\ensuremath{\mbox{Cov}}}

\begin{document}    

\title[Removal of zero-point drift]{Removal of zero-point drift  from AB data and the statistical cost}
\author{H Erik Swanson and Stephan Schlamminger}
\address{Center for Experimental Nuclear and Particle Astrophysics, University of Washington, Seattle, Washington 98195 USA}
\eads{eriks@u.washington.edu,schlammi@u.washington.edu}
\submitto{\MST}

\begin{abstract}
Often the result of a scientific experiment is given by the difference of measurements in two configurations, denoted by $A$ and $B$. Since the measurements are not obtained simultaneously, drift of the zero-point can bias the result. In practice measurement patterns are used to minimize this bias. The time sequence $AB$ followed by $BA$, for example, would cancel a linear drift in the average difference $A-B$. We propose taking data with an alternating series $ABAB$.., and removing drift with a post-hoc analysis. We present an analysis method that removes bias from the result for drift up to polynomial order $p$. A statistical cost function $c(N)$ is introduced to compare the uncertainty in the end result with that from using a raw data average. For a data set size $N>30$ the statistical cost is negligible. For $N<30$ the cost is plotted as a function of $N$ and filter order $p$ and the trade off between the size of the data set and $p$ is discussed.
\end{abstract}

\section{Motivation and introduction}
\label{sec:Mot}
The accuracy and precision of a measurement are reduced by drifts in the  offset and sensitivity of the instrument used to collect the data. The offset or zero-point  is additive  and  independent of the magnitude of the value being measured. If uncorrected,  it directly biases the result of the measurement. Sensitivity on the other hand is multiplicative and  affects the output of the instrument more for larger quantities than for smaller ones. Sensitivity can be measured and any variation corrected by repeated or even continuous calibration of the instrument. The offset can be eliminated  using modulation, i.e. switching the measurement setup such that the quantity being measured alternates in sign with  measurement states $A$ and $B$.  As long as  the zero-point does not change  over the time it takes to make the measurement,  the result determined from the difference of these states   is independent of the offset. As an example, a  technique  employed by many laboratories  uses a single pan balance  for making mass measurements. The sample $A$ and a known calibrated mass  $B$ are alternately placed on the balance's pan and their values read out. The mass difference, obtained by subtracting two subsequent readings, and the known mass $B$ give the mass of $A$. An offset that is constant in time does not bias the mass difference. The result can be made less dependent on the sensitivity of the balance by choosing a calibration weight whose mass is nearly identical to that of sample. Alternately, this can be viewed as modulation where the measurement result is half the difference of the measured masses and the zero-point is their average. 

If the time required to make an $AB$ comparison is sufficiently long compared to the drift rate, the assumption of an unchanging zero-point  is no longer  valid. In this case the result includes a  systematic bias from the zero-point drift. This problem is more acute for measurements taken over very long times due to the $1/f$ behavior of noise. The term $1/f$ noise reflects the fact that the power spectral density of the noise  is inversely proportional to the frequency~\cite{Milotti02}. Measurements made over long time intervals would experience larger fluctuations in the zero-point than those over short time intervals. As a consequence, an ensemble of mean values taken at different times has more scatter than the errors of the means themselves. The presence of $1/f$ noise has been shown to be fully consistent with a stationary distribution~\cite{stoisiek76} and this extra noise can be added to the standard deviation of the distribution.

The technique described in the above example is not limited to mass metrology. Many precision experiments change one critical control parameter or a part of the geometry to achieve modulation in the experiment's output. Examples include measurements of the gravitational constant~\cite{Sch06}, the search for parity violation in nucleon-nucleon interactions~\cite{Swa09,Bass09}, the determination of Planck's constant using a Watt balance~\cite{Ste05} and searches for equivalence-principle violations~\cite{Sch08}. As the examples indicate square wave modulation is a powerful tool used for Null and non-Null experiments. The lock-in amplifier is able to recover noisy signals using similar techniques. 

Consider a series of measurements,${u_i}$, spaced equally in time  where an observable $\eta$ changes sign with each measurement, $z_0(i)$ denotes the zero-point that varies slowly with $i$, and $\varepsilon_i$ is a random variable that adds noise to the measurement. The expectation value of $\varepsilon$ is $E(\varepsilon)=0$ and the noise is assumed to be uncorrelated and stationary, $\Cov{(\varepsilon_i,\varepsilon_j)}=\delta_{ij} \sigma^2$.  The time series can be written as
\begin{equation}
u_i = z_0(i) + (-1)^i \eta + \varepsilon_i,
\label{eqn:def1}
\end{equation}
where $i$ runs from 1 to $N$, which is assumed to be even.
An estimator $S$ of the signal can be obtained by multiplying the time series by $(-1)^i$ and subsequent averaging,
\begin{equation}
S = \frac{1}{N} \sum_i ^N \left( (-1)^i z_0(i)  + \eta + (-1)^i \varepsilon_i \right).
\end{equation}
The expectation value and the variance of the estimator are 
\begin{eqnarray}
E(S) &=& \eta + \frac{1}{2} E (z_0\prime ) \\
\Var(S) &=& E (z_0 \prime^2) -  (E {z_0}\prime)^2 +\sigma^2
\end{eqnarray}
The signal $\eta$ is recovered, however it is biased by the mean value of the slope of the varying background, $z_0\prime\equiv z_0(i+1)-z_0(i)$. The variance of the estimator contains the variance of the measurement noise $\sigma^2$ plus variations about the mean slope of $z_0(i)$. This later part limits the ultimate statistical precision of the measurement.

Various patterns for taking data are known to completely cancel linear and even higher order drifts in calculations of the mean~\cite{Sut93,Gla00}.  For example, the sequences $ABA$ and $ABBA$ both remove linear drift terms.
$ABBA$  with 4 measurements ${u_1,u_2,u_3,u_4}$, recovers the signal by computing the mean according to the expression  $S_4 = \frac{1}{4} (u_1-u_2 - (u_3-u_4))$. Assuming uncorrelated data ($\Cov{(u_i,u_j)}=\delta_{ij} \sigma^2$), the variance of this estimator is given by $\Var{(S_4)}=\sigma^2/4$.  Averaging $N/4$ of these $S_4$ results gives an overall variance of  $\sigma^2/N$. 
In the $ABA$ sequence, the estimator is given by $S_3 = \frac{1}{4} (u_1 -2u_2 + u_3)$ with $\Var{(S_3)}=3\sigma^2/8$. Averaging $N/3$ of these $S_3$ results leads to a variance of  $(9/8)\sigma^2/N$.
In general, the smallest possible variance is achieved when all measurements are weighted with equal absolute values. When comparing different drift cancellation methods a useful figure of merit is the relative increase in the uncertainty of the result $S_k$ over the minimum variance $\sigma/\sqrt{N}$. Thus  we define the statistical cost of $S_k$ as
\begin{equation*}
c(N) \equiv  \sqrt{ \frac{N} {\sigma^2 }\Var(S_k) } -1.
\nonumber 
\end{equation*}
Comparing the statistical cost values, 0.06 for $ABA$ and 0 for $ABBA$, shows the latter to be better. Strictly speaking, the data set size $N$ should be divisible by both $3$ and $4$ when directly comparing these two methods.  Although by this metric $S_4$ is the best unbiased estimator in the presence of linear drift, the $ABBA$ pattern has  several disadvantages: (1) The series is asymmetric. The first $B$ occurs after a configuration change, whereas no such change occurs for the second $B$ measurement. It is reasonable to assume the two $B$ measurements have different uncertainties and drift properties.   (2) The measurement pattern is fixed in advance. If the data exhibits additional non-linear drift, removing it would require a new measurement with a more appropriate pattern. (3) Drift cancellation only occurs if all measurements of the series are used. If one measurement is missing or affected by an out-lier, the remaining measurements in that sequence would have to be discarded.

In this article we propose an alternative approach  that avoids the above shortcomings. Data is taken in an alternating (square wave) sequence $ABAB...$ without regard for any inherent drift. An estimator $F_N^p$  is then constructed that removes drift up to any desired polynomial order $p$ where $p<N$. When applied to the data it yields an expectation value for the mean independent of the  drift orders selected. This plays the same roll as the data taking sequences discussed above where linear drift is removed.   The statistical cost associated with $F_N^p$ increases with increasing $p$ and decreases with increasing $N$.   A cost -- benefit analysis can help choose  appropriate values for $p$ and $N$.

The next two sections contain the mathematical justification for the methods used. The construction of the estimator $F_N^p$ is described in two steps. First, we introduce a filter that removes  polynomial drift from the input data. The output of this filter is a shorter time series without drift to the order  selected. Next, the filtered data are averaged to a single number which is  the best estimate of the signal. In practice, the estimator $F_N^p$ is used to generate a set of weights that are  applied to the entire data set. These products are then summed together for the result.

\section{Design of a filter that removes polynomial drift from data}
\label{sec:filt_design}

The filter is a linear combination of consecutive terms in the original data series ${u_1,u_2,..,u_N}$. A new series ${y_1,y_2,...,y_{N-p-1}}$ is calculated as
\begin{equation}
y_i = \sum_{k=0}^{p} C_k u_{i+k}
\label{eq:lin-comb}
\end{equation}
The $C_k$s are chosen to cancel variations across the included points that are less than a given polynomial order $p$ and  preserve a signal that changes sign with each successive term.  
 These design requirements for the filter can be met by solving the matrix equation
\begin{eqnarray}
 \left(
\begin{array}{ccccccc}
1  	&1 	&1     	& 1 	&\cdots	 &1  \\
0 	&1  	&2   	& 3 	&\cdots	 &p  \\
0    	&1 	&4   	& 9 	&\cdots	 &p^2 \\
 	&   	& 	&\vdots	&	 &   \\
0   & 1    	& 2^{p-1} & 3^{p-1} &\cdots	 &p^{p-1} \\
1 	&-1   	&1 	&-1	&\cdots	 &(-1)^p \\
\end{array} 
\right)
 \left(
\begin{array}{c}
C_0 \\ C_1 \\ C_2 \\ \vdots \\ C_{p-1} \\ C_p \\
\end{array} 
\right)=
\left(
\begin{array}{c}
0 \\ 0 \\ 0 \\ \vdots \\ 0 \\1 \\
\end{array} 
\right),
\label{eq:matrix}
\end{eqnarray}
for $C_k$. The first $p$ rows cancel variations with polynomial orders up to $p-1$ including the zeroth order, i.e. offset. The last row normalizes the part of $u_i$ that changes sign with each successive $i$. The $p+1$ equations uniquely determine $p+1$ coefficients of the sum in equation~(\ref{eq:lin-comb}).

An alternate analytical expression for these coefficients can be obtained from the generating function $F(z) = (1-z^{-1})^{p}$ where $z$ is complex. In the limit $|z|\rightarrow 1$ it can be seen that the derivatives of $F(z)$ less than order $p$ vanish. This has a correspondence with equation~(\ref{eq:matrix}), where in the first $p$ rows the sum of each successive polynomial order is required to vanish. The binomial expansion of $F(z)$ gives

\begin{equation}
F(z) = (1-z^{-1})^p = \sum_{k=0}^p (-1)^k \left( \begin{array}{c}
p \\ k
\end{array} \right) z^{-k} \label{eqn:zt}
\end{equation}
$F(z)$  is therefore the Z-transform of a discrete time series $f[k]$ with the property that the sum of terms $k^jf[k]$ equal zero for $j < p$. By taking the Z-transform of both sides of equation~(\ref{eq:lin-comb}) its transfer function $H(z)$ can be computed, 

\begin{equation}
H(z) = \frac{Y(z)}{U(z)} = \sum_{k=0}^p C_k z^{-k}.
\end{equation}

Equating $H(z)$ to $F(z)$ and normalizing with $1/2^p$ to preserve the signal amplitude, the coefficients
\begin{equation}
C_k = \frac{(-1)^k}{2^p}
\left( \begin{array}{c}
p \\ k
\end{array} \right)
\end{equation}
are obtained.

Substituting these coefficients into equation~(\ref{eq:lin-comb}) one obtains
\begin{equation}
y_i(p) = \frac{1}{2^p}\sum_{k=0}^p  (-1)^{i+k}  \left( {p \atop k} \right) u_{i+k}. 
\label{eq:lin}
\end{equation}
An additional factor of $(-1)^{i}$ is included here to demodulate the alternating signal $\eta$ found in equation~(\ref{eqn:def1}).

Equation~(\ref{eq:lin}) is a discrete convolution of the input data $u_i$ with a finite input response (FIR) filter kernel. The filter described here has the property of removing polynomial drifts. In principle, any (FIR) filter could be used, and another  may be more efficient in environments where the functional form of the drift is known.

To properly account for uncertainties it is advantageous to express the convolution as a linear transformation of the input data set $\bi{U}={u_1,..,u_N}$ to a filtered data set $\bi{Y}={y_1,..,y_{N-p}}$:
\begin{equation}
\bi{Y}=\bi{A}\bi{U}
\label{eq:trans}
\end{equation}
The matrix $\bi{A}$ has dimensions $(N-p) \times N$ as incompletely defined terms in the convolution are discarded. The elements of each row are taken from the coefficients in equation~(\ref{eq:lin}):
\begin{equation}
a_{ij} = \frac{1}{2^p} \left( {p \atop j-i} \right)(-1)^{j}  \mbox{~~~for $0\le j-i \le  p$, zero otherwise}.
\label{eq:aij}
\end{equation}
Adjacent values in the filtered data set are not statistically independent, since filtering introduces correlations between the elements of $\bi{Y}$. The covariance matrix of the filtered data can be calculated from the covariance matrix of the input data $ \Cov{(U,U)} $ using,
\begin{equation}
\bi{C} \equiv \Cov{(Y,Y)}  =\bi{A} \Cov{(U,U)} \bi{A}^T.
\end{equation}
We make the assumption that the $u_i$ values stem from an uncorrelated and stationary distribution with sample variance $\sigma^2$ (See the $1/f$ noise discussion in section~\ref{sec:Mot}).  The covariance matrix $\Cov{(U,U)}$ can then be written as the $N\times N$ identity matrix scaled by $\sigma^2$, however, the calculations presented here do not require a priori knowledge of $\sigma$. The elements of $\bi{C}$ are given by,
\begin{eqnarray}
c_{ij} &=& \frac{\sigma^2}{4^p}\sum_{r=0}^{N-p} a_{ir} a_{jr} \\
 &=& \frac{\sigma^2}{4^p}\sum_{r=0}^{N-p}(-1)^{2r}\left( {p \atop r-j} \right)  \left( {p \atop r-i} \right),
\label{eq:AAT1}
\end{eqnarray}
which simplifies using Vandermonde's identity and a change of variables to
\begin{equation}
c_{ij} = \frac{\sigma^2}{4^p}\left( {2p \atop p+i-j} \right) \mbox{~~~for $-p\le i-j \le  p$, zero otherwise}. 
\label{eq:AAT2}
\end{equation}
Every covariance matrix is positive semi definite and symmetric, i.e. $\bi{C} = \bi{C}^T$~\cite{Strang}. Here, the latter is apparent from the inherent symmetry of the binomial coefficients. The covariance matrix $\bi{C}$ is no longer diagonal and the degree of correlation between the filtered data points is given by the off-diagonal elements.

\section{Extracting the signal from drift-free data}
\label{sec:drift_removal}

The filtered data set, $\bi{Y}$, contains the demodulated difference signal $\eta$, which can be retrieved using linear regression to estimate its mean value $\mu$. Given the form of the covariance matrix, the best linear unbiased estimator (BLUE) is obtained using Aitken's generalized least squares procedure~\cite{Aitken35, Rao}. 

\begin{equation}
\mu = \frac{\bi{X}^T {\bi C}^{-1} } {\bi{X}^T  {\bi C}^{-1} \bi{X}} \bi{Y},
\end{equation}
where the design matrix $\bi{X}$, is a column vector of length $N-p$ with all elements equal to one. It should be noted that the matrix expressions in the denominators evaluate to scalar values. For notational purposes, we define the estimator of our signal $\mu$ to be $F_N^p$ in order to reflect its two parameter nature ($N$ data points and drift reduction up to order $p$). Since the result is independent of the data uncertainty $\sigma^2$, it can be expressed in terms of known quantities,
\begin{equation}
F_N^p= \frac{\bi{X}^T (\bi{A}\bi{A}^T )^{-1} } {\bi{X}^T  (\bi{A}\bi{A}^T )^{-1} \bi{X}} \bi{Y}. 
\label{eqn:estm}
\end{equation}
Replacing $\bi{Y}$ by $\bi{AU}$, a weight vector $\bi{V}$ can be defined which allows calculation of the result directly from the input data set.
The estimate for the signal can be obtained in one simple step, the scalar product,
\begin{equation}
F_N^p = \bi{V} \bi{U} \;\;\;\mbox{with}\;\;\;\bi{V} =  \frac{\bi{X}^T (\bi{A}\bi{A}^T )^{-1} } {\bi{X}^T  (\bi{A}\bi{A}^T )^{-1} \bi{X}} \bi{A}.
\end{equation}

The weight vector $\bi{V}$ is unique to the values of $N$ and $p$ but can be reused as long as the length of the data set and the desired order of drift suppression remain unchanged.

\section{The variance of the estimator}
\label{sec:variance}

The variance of $F_N^p$ is calculated using error propagation. Let the factor multiplying $\bi{Y}$ in equation~(\ref{eqn:estm}) be
\begin{equation}
\bi{W} = \frac{\bi{X}^T (\bi{A}\bi{A}^T )^{-1} } {\bi{X}^T  (\bi{A}\bi{A}^T )^{-1} \bi{X}}.
\label{eqn:w}
\end{equation}
It then follows that the variance of $F_N^p$ is given by
\begin{equation*}
\Var(F_N^p) = \bi{W} \Cov{(Y,Y)} \bi{W}^T,
\end{equation*}
which simplifies to
\begin{equation}
\Var(F_N^p) = \frac{\sigma^2}{\bi{X}^T  (\bi{A}\bi{A}^T )^{-1} \bi{X}}  \label{eq:variance_wm}.
\end{equation}
The variance of the estimator is the inverse of the sum of all elements of the inverse correlation matrix scaled by $\sigma^2$.
If the value of $\sigma^2$ is not known, one estimate for $\Var(F_N^p)$ can be obtained by calculating the scatter in the estimator for a number of different data segments each of length $N$. If only a single set of $N$ measurements is available, $\sigma^2$ can be estimated from the sum of squares of the residuals about the regression mean $\mu$ as shown.
The quantity
\begin{equation*}
(\bi{Y}-\mu \bi{X})^T \bi{C}^{-1}(\bi{Y}-\mu \bi{X})
\nonumber
\end{equation*}
follows a $\chi^2$-distribution which has a mean value equal to the number of degrees of freedom $\nu = N - p -1$. Hence, 
\begin{equation}
s^2 = (\bi{Y}-\mu \bi{X})^T (\bi{A}\bi{A}^T )^{-1}(\bi{Y}-\mu \bi{X})/(N-p-1) \label{eq:s}
\end{equation}
is an estimator of $\sigma^2$.
Using equation~(\ref{eq:variance_wm})  the variance in the mean of the filtered data can be compared to that of the un-filtered input data.  The corresponding variance in the mean of $N$ un-correlated data points is $\sigma^2 N^{-1}$  where the data variability $\sigma^2$ is common to both. Using the ratio of their variances, the statistical cost function becomes
\begin{equation}
c(N) =  \sqrt{\frac{N}{\bi{X}^T  (\bi{A}\bi{A}^T )^{-1} \bi{X}}}-1 .
\label{eq:cost}
\end{equation}
This gives the relative increase in statistical error resulting from the smaller correlated data set $(N-p)$ obtained in equation~(\ref{eq:trans}).

\begin{figure}[ht!]
\begin{center}

\includegraphics[height=8cm, clip]{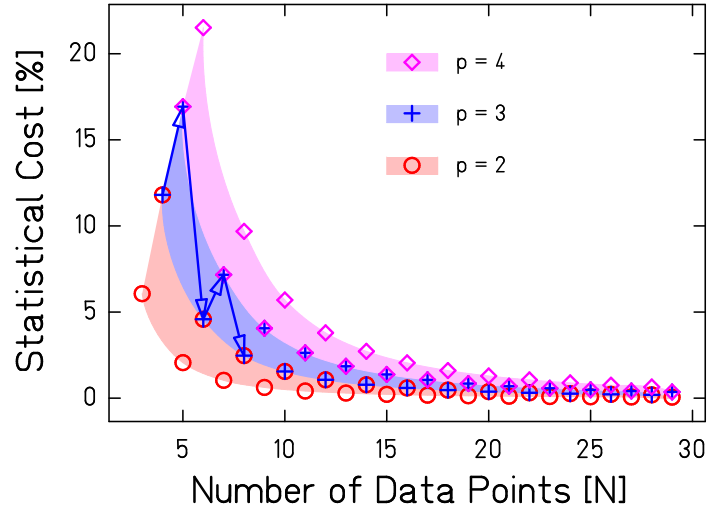}
\end{center}
\caption{The statistical cost is shown as a percentage versus the number of data points. The shaded bands are included to guide the eye and correspond to three different order filters. The cost values for a given filter order $p$ oscillate between the margins of their respective band with increasing $N$. This trajectory is shown using arrows for the first few points of the $p = 3$ filter.}
\label{fig:weightedmean}
\end{figure}

Figure \ref{fig:weightedmean} shows the statistical cost for three different filters labeled by their order $p$ versus the number of data points $N$. Cost values for a fixed value of $p$ oscillate between the upper and lower edges of their respective shaded bands with increasing $N$. The upper cost value for a filter of order $p$ is the same as the lower value for a filter with $p+1$. 

This oscillation can be understood as follows:
Each $y_i$ is a particular linear combination of the raw data as given in equation~(\ref{eq:lin}). If they have the same weighting factor, we can sum two at a time where their $i$ values differ by $m$.
\begin{equation}
y_i(p)  + y_{i+m}(p) = \frac{1}{2^p}\sum_{k=0}^p  (-1)^{i+k}  \left( {p \atop k} \right)( u_{i+k} + (-1)^m u_{i+k+m}). 
\end{equation}
If $m$ is odd this is equivalent to applying the filter to the difference of the two points. The filter removes variations across the points that go as $i^q$ where $q$ is some power less than $p$. The difference $(i+m)^q - i^q$ is always one order less than $q$. In the process of taking the mean, for those points whose index differs by an odd number, the data is effectively filtered to the next higher order than the one chosen. In computing the mean, the  $y_i$ values are weighted by the vector $\bi{W}$ given in equation~(\ref{eqn:w}). Elements of this weight vector are each a sum over the corresponding column of the inverse correlation matrix $(\bi{A}\bi{A}^T )^{-1}$. Since the correlation matrix is symmetric about its diagonal it follows that the inverse correlation matrix is symmetric as well.\footnote{This follows from a property of invertible matrices $(\bi{C^{-1}})^T = (\bi{C^T})^{-1}$}  The weight vector therefore has this same symmetry about its center. If the length of the weight vector  is even, then the positions of its elements that are equal  in value differ by an odd number and the polynomial order filtered increases by 1. The oscillating behavior in the variance reflects this additional cancellation. When choosing either the filter order or the length of the data set, the statistical error is smallest when $N-p$ is odd.

The statistical precision of the mean of a measurement series scales as $\sigma/\sqrt{N}$ where $N$ is the number of measurements. If all measurements come from a population with variance  known a priori to be small enough, the desired statistical precision can be achieved with just a few measurements. In this case drift suppression is best achieved with sequence patterns as discussed in section~\ref{sec:Mot}.  In all other cases  precision measurements  require $N$  sufficiently large that the increase in  error from the filter would be minimal.
In  measurement scenarios with two different outcomes, equal amounts of data are taken in each state to insure  an unbiased result. The lowest order filter consistent with even values of $N$ has $p = 3$. It can be seen from figure~\ref{fig:weightedmean} that after  8  measurements the increase in statistical error is below a few percent. If only  linear drift  were removed, this same  statistical cost  could be achieved  with even fewer measurement. If 20 or more measurements are available the increase in error is minimal for all filter orders shown.

This analysis method can also be used to characterize the zero-point drift of the instrument. We start with the null hypothesis, i.e. that no significant drift exists so the results would be the same whether or not the filter is used. Without loss in generality we assume an even $N$. Hence, we choose to compare this with a filter of order $p = 3$ to achieve the smallest error.  An estimator for un-filtered data can be obtained by letting $p \rightarrow 0$. In this case elements of the $\bi{A}$ matrix ($N \times N $ ) are given by $a_{i,j} = (-1)^i\delta _{i,j}$ and the design matrix $\bi{X}$ has length $N$. Since both the filtered and un-filtered estimators are calculated from the same raw data, the null hypothesis can be checked. Using equation~(\ref{eq:s}) an estimator of the standard deviation for both cases can be obtained, $s^2_{p=0}$ and $s^2_{p=3}$. Multiplying these estimators  by their degrees of freedom $\nu = N-p-1$, yields variables that follow  $\chi^2$-distributions with $N-1$ and $N-4$ degrees of freedom respectively. In each case the $\chi^2$ probability is calculated and the null hypothesis is accepted if the $\chi^2$ probability of the filtered data is not significantly larger than that of the un-filtered data.

\section{Conclusion}

We have shown how zero-point drift can introduce a bias into the outcome of a measurement. Difference measurements in particular are  sensitive to drifts, especially when the time required for a measurement is long. If there are two states $A$ and $B$, choosing specific measurement sequences like $ABBA$ can completely eliminate this  bias from the difference. 

The alternative method proposed here acquires data with a square wave modulation sequence $ABAB$... and recovers the signal with a post-hoc analysis. It is equally effective in removing linear drift and, in addition, can be configured to suppress zero-point drift of any desired polynomial order. Besides the simplicity of the measurement sequence, it offers the following advantages: (1) The measurement series is symmetric as every measurement is performed after the state has changed, (2) the highest order drift  suppressed can be selected after the data has been successfully taken, and (3) a missing data point has only a minor effect on  drift suppression. Its main disadvantage  is that it results in a slightly higher uncertainty in the end result when compared with schemes like $ABBA$ where data points are weighted with equal absolute values. The relative amount by which the uncertainty is increased depends on the number of data points $N$ and the order of the polynomial drift suppressed ($<p$) but is typically less than a few percent as shown in figure~\ref{fig:weightedmean}. 

In the analysis process the set of weights (weight vector) used to average the $ABAB$... data series is calculated according to the prescription given for the estimator $F_N^p$. 
Applying the method is straight forward due to the use of matrix equations. Typically the filter would be chosen to have an order $p$ of 2 or 3 depending on whether the data set size $N$ is odd or even, respectively. With $p$ and $N$ known one need only construct the appropriate $\bi{A}$ matrix and  design matrix $\bi{X}$. 
There are software packages available such as MatLab, Mathematica, Wavemetric's IGOR, and Python with SciPi and NumPi to perform the matrix operations. The estimator  of the result is then the scalar product of the data vector with the weight vector.

We have chosen a polynomial expansion for the drift because the lowest order term contributes directly as a bias offset of the measurement. There may be specific environments where a different expansion would lead to smaller errors with fewer terms.  
The matrix equations have been kept sufficiently general as to allow incorporating other functional forms of the drift. We note that the formalism presented here can be adopted to accommodate irregularly-spaced data. In this case each row of matrix, $\bf{A}$, has to be changed to meet the irregularly-spaced analog of equation~(\ref{eq:matrix}).

In conclusion, we believe that the filter presented here in combination with an $ABAB$... data sequence is an elegant and powerful tool for obtaining accurate measurements in the presence of drift.

\section*{References}
\bibliographystyle{unsrt}
\bibliography{aba}

\appendix
\section{An example calculation}
In this example we choose a filter of order $p = 2$ as 
 a closed form expression for the weight vector $\bi{V}$ for odd $N$ can be written as
\begin{equation}
\bi{V} = \left( \frac{1}{N+1}, \frac{-1}{N-1}, \frac{1}{N+1}, \frac{-1}{N-1}, \dots \frac{1}{N+1} \right) \textnormal{~~~for~~} N \ge 3.
\end{equation}
The estimator of the signal is given by the dot product of $\bi{V}$ with the data vector $\bi{U}$. The reader is encouraged to calculate the mean with various data sets of his or her choosing. For example:
\begin{eqnarray}
\bi{U}^T = (1,1,1,1,\dots1) &\;\;\mbox{constant offset}\;\;& F_N^2=\bi{V}\bi{U}=0 \nonumber\\
\bi{U}^T = (1,2,3,4,\dots N) &\;\;\mbox{linear drift}\;\;& F_N^2=\bi{V}\bi{U}=0 \nonumber\\
\bi{U}^T = (1,4,9,16,\dots N^2)&\;\;\mbox{quadratic drift}\;\;&F_N^2=\bi{V}\bi{U}\ne0 \nonumber \\
\bi{U}^T = (d,-d,d,-d\dots d) &\;\;\mbox{signal}\;\;& F_N^2=\bi{V}\bi{U}=d \nonumber
\end{eqnarray}
The variance of the estimator is given by the length of the weight vector squared times the variance of the original data, $\Var(F_N^p)=\bi{V}\bi{V}^T\sigma^2$. In this example, $p=2$ and  odd $N$, the variance calculates to $\Var(F_N^p)=\sigma^2 N/(N^2-1)$, which should be compared with $\sigma^2/N$, the smallest possible variance for an estimator of the signal.  The reader is encouraged to add Gaussian noise to the data examples above to numerically verify the variance of the estimator.

An interval of $1/f$ noise can be expanded in a polynomial series and this result shows terms above the filter cutoff  bias the mean value. An ensemble of mean values taken at different times will tend to scatter more than the errors of the means themselves.

\end{document}